\documentclass[aps,prl,twocolumn,superscriptaddress,showpacs]{revtex4}
\usepackage{graphicx}

\begin{document}

\title{Complete Electric Dipole Response and the Neutron Skin in $^{208}$Pb}

\newcommand{\RCNP}{Research Center for Nuclear Physics, Osaka University, Ibaraki, Osaka 567-0047, Japan}
\newcommand{\Wit}{School of Physics, University of the Witwatersrand, Johannesburg 2050, South Africa}
\newcommand{\Kyushu}{Department of Physics, Kyushu University, Fukuoka 812-8581, Japan}
\newcommand{\iThemba}{iThemba LABS, Somerset West 7129, South Africa}
\newcommand{\Osaka}{Department of Physics, Osaka University, Toyonaka, Osaka 560-0043, Japan}
\newcommand{\CYRIC}{Cyclotron and Radioisotope Center, Tohoku University, Sendai,  980-8578, Japan}
\newcommand{\CNS}{Center for Nuclear Study, University of Tokyo, Bunkyo,  Tokyo 113-0033, Japan}
\newcommand{\TUDarmstadt}{Institut f\"ur Kernphysik, Technische Universit\"{a}t Darmstadt, D-64289 Darmstadt, Germany}
\newcommand{\Valencia}{Instituto de Fisica Corpuscular, CSIC-Universidad de Valencia, E-46071 Valencia, Spain}
\newcommand{\MSU}{NSCL, Michigan State Univ., MI 48824, USA}
\newcommand{\Kyoto}{Department of Physics, Kyoto University, Kyoto 606-8502, Japan}
\newcommand{\Niigata}{Department of Physics, Niigata University, Niigata 950-2102, Japan}
\newcommand{\RIKEN}{RIKEN Nishina Center, Wako, Saitama 351-0198, Japan}
\newcommand{\HIMAC}{National Institute of Radiological Sciences, Chiba 263-8555, Japan}
\newcommand{\ECT}{ECT*, Villa Tambosi, I-38123, Villazzano (Trento), Italy}
\newcommand{\KVI}{Kernfysisch Versneller Instituut, University of Groningen, Zernikelaan 25, NL-9747 AA Groningen, The Netherlands}
\newcommand{\TexasAM}{Department of Physics and Astronomy, Texas A\&M University-Commerce, Commerce, Texas 75429, USA}
\newcommand{\GSI}{GSI Helmholtzzentrum f\"{u}r Schwerionenforschung, D-64291 Darmstadt, Germany}
\newcommand{\Frankfurt}{Institut f\"{u}r Theoretische Physik,
Goethe-Universit\"{a}t, 60438 Frankfurt am Main, Germany}

\author{A.~Tamii}\affiliation{\RCNP}
%\email[]{tamii@rcnp.oaska-u.ac.jp}\affiliation{\RCNP}
\author{I.~Poltoratska}\affiliation{\TUDarmstadt}
\author{P.~von~Neumann-Cosel}\email{vnc@ikp.tu-darmstadt.de} \affiliation{\TUDarmstadt}
\author{Y.~Fujita}\affiliation{\Osaka}
\author{T.~Adachi}\affiliation{\Osaka} \affiliation{\KVI}
\author{C.~A.~Bertulani}\affiliation{\TexasAM}
\author{J.~Carter}\affiliation{\Wit}
\author{M.~Dozono}\affiliation{\Kyushu}
\author{H.~Fujita}\affiliation{\RCNP}
\author{K.~Fujita}\affiliation{\Kyushu}
\author{K.~Hatanaka}\affiliation{\RCNP}
\author{A.~M.~Heilmann}\affiliation{\TUDarmstadt}
\author{D.~Ishikawa}\affiliation{\RCNP}
\author{M.~Itoh}\affiliation{\CYRIC}
\author{T.~Kawabata}\affiliation{\Kyoto}
\author{Y.~Kalmykov}\affiliation{\TUDarmstadt}
\author{E.~Litvinova}\affiliation{\GSI} \affiliation{\Frankfurt}
\author{H.~Matsubara}\affiliation{\CNS}
\author{K.~Nakanishi}\affiliation{\CNS}
\author{R.~Neveling}\affiliation{\iThemba}
\author{H.~Okamura}\affiliation{\RCNP}
\author{H.~J.~Ong}\affiliation{\RCNP}
\author{B.~\"{Ozel-Tashenov}}\affiliation{\GSI}
\author{V.~Yu.~Ponomarev}\affiliation{\TUDarmstadt}
\author{A.~Richter}\affiliation{\TUDarmstadt}\affiliation{\ECT}
\author{B.~Rubio}\affiliation{\Valencia}
\author{H.~Sakaguchi}\affiliation{\RCNP}
\author{Y.~Sakemi}\affiliation{\CYRIC}
\author{Y.~Sasamoto}\affiliation{\CNS}
\author{Y.~Shimbara} \affiliation{\Osaka} \affiliation{\Niigata}
\author{Y.~Shimizu}\affiliation{\RIKEN}
\author{F.~D.~Smit}\affiliation{\iThemba}
\author{T.~Suzuki}\affiliation{\RCNP}
\author{Y.~Tameshige}\affiliation{\HIMAC}
\author{J.~Wambach}\affiliation{\TUDarmstadt}
\author{R.~Yamada}\affiliation{\Niigata}
\author{M.~Yosoi}\affiliation{\RCNP}
\author{J.~Zenihiro}\affiliation{\RCNP}

\date{\today}

\begin{abstract}
 A benchmark experiment on $^{208}$Pb shows that polarized proton inelastic scattering at very forward angles including $0^\circ$ is a powerful tool for high-resolution studies of electric dipole ($E1$) and spin magnetic dipole ($M1$) modes in nuclei over a broad excitation energy range to test up-to-date nuclear models. The extracted $E1$ polarizability leads to a neutron skin thickness $r_{\rm skin} = 0.156^{+0.025}_{-0.021}$~fm in $^{208}$Pb derived within a mean-field model [Phys.~Rev.~C {\bf 81}, 051303 (2010)], thereby constraining the symmetry energy and its density dependence, relevant to the description of neutron stars.
\end{abstract}


\maketitle

The electric dipole ($E1$) response of nuclei is dominated by the giant dipole resonance (GDR), a highly excited collective mode above the particle emission threshold \cite{ber75}. Its properties are  well understood but recent interest focusses on evidence for a soft mode in neutron-rich nuclei below the GDR termed pygmy dipole resonance (PDR). Because of the saturation of nuclear density, excess neutrons might form a skin whose oscillations against an isospin-saturated core should give rise to a low-energy $E1$ mode \cite{paa07}. Therefore, the PDR may shed light onto the formation of neutron skins in nuclei \cite{pie06}. Another quantity related to nuclear $E1$ modes is the symmetry energy acting as restoring force. The $E1$ strength distribution carries information on its poorly known magnitude and density dependence \cite{car10}, indispensable ingredients for the modeling of the equilibrium properties of neutron stars \cite{hor01}.

A case of special interest is the doubly magic nucleus $^{208}$Pb. In a measurement of parity-violating elastic electron scattering  at  JLAB, the PREX collaboration \cite{prex} aimed at the first model-independent determination of the neutron skin thickness in $^{208}$Pb. However, the recent result $r_{\rm skin} = 0.34^{+0.15}_{-0.17}$~fm suffers still from limited statistics. Studies of energy density functionals (EDFs) \cite{ben03} using Skyrme forces \cite{rei10} or a relativistic framework \cite{pie10} suggest the nuclear dipole polarizability $\alpha_D$ as an alternative observable constraining both neutron skin and symmetry energy. The polarizability is related to the photoabsorption cross section $\sigma_{abs}$ \cite{boh81}
\begin{equation}
 \alpha_D = \frac{\hbar c}{2 \pi^2 e^2} %\sum \!\!\!\!\!\!\!\!
 \int \frac{\sigma_{abs}}{\omega^2} \, d\omega,
 \label{eq:pol}
 \end{equation}
where $\omega$ denotes the photon energy. Because of the inverse energy weighting, $\alpha_D$ depends on the $E1$ strength at low energies.
% Shortened 24.5.11
Theoretically, advanced methods exist in closed-shell nuclei for a realistic description of the $E1$ strength distributions \cite{spe91}.

The centroid of the PDR lies typically in the vicinity of the neutron emission threshold ($S_n$). Data on the PDR in very neutron-rich nuclei are still scarce \cite{adr05,kli07,wie09}. Stable nuclei at different shell closures have been explored  with the $(\gamma,\gamma')$ reaction (Ref.~\cite{sav08} and refs.\ therein). While this technique provides important information on the fine structure of the PDR, it is essentially limited to excitation energies up to $S_n$, and unobserved branching ratios of the $\gamma$ decay to excited states may require corrections for the total strength \cite{rus08}. Measurements of decay neutrons are constrained to energies $E_x > S_n$ and uncertainties in the vicinity of $S_n$ are large. We present here a new experimental tool, viz.\ polarized proton scattering at angles close to and including $0^\circ$, to provide the {\em complete E1 response in nuclei} up to excitation energies well above the region of the GDR. At proton energies of $200-400$ MeV the cross sections at small momentum transfers are dominated by isovector spinflip-$M1$ transitions (the analog of the Gamow-Teller mode) and by Coulomb excitation of non-spinflip $E1$ transitions \cite{fre90,hey10}.  A separation of these two contributions, necessary for an extraction of the $E1$ response, is achieved by two independent methods: a multipole decomposition analysis  of the angular distributions (MDA) and the measurement of polarization transfer (PT) observables.

The $^{208}$Pb($\vec{p},\vec{p}'$) experiment was performed at RCNP, Osaka, Japan. Details of the technique can be found in \cite{tam09}.
In the present work \cite{pol11}, a proton beam of 295 MeV with intensities  $2-10$~nA and an average polarization  $P_0 \simeq 0.7$ bombarded an isotopically enriched $^{208}$Pb foil with an areal density of 5.2 mg/cm$^2$.
Data were taken with the Grand Raiden spectrometer \cite{fuj99} in an angular range $0^\circ - 2.5^\circ$ and for excitation energies $E_x \simeq 5 - 23$ MeV.
%The excitation energy resolution was $25 - 30$~keV (full width at half %maximum, FWHM).
Sideway ($S$) and longitudinally ($L$) polarized proton beams were used to measure the
polarization transfer coefficients \cite{ohl79} $D_{SS'}$ and $D_{LL'}$, respectively.
Additional data with unpolarized protons were taken at angles up to $10^\circ$.
Utilizing dispersion matching techniques, a high energy resolution $\Delta E \simeq 25$ keV (full width at half maximum) could be achieved.
\begin{figure}[tbh]
\includegraphics[width=8cm]{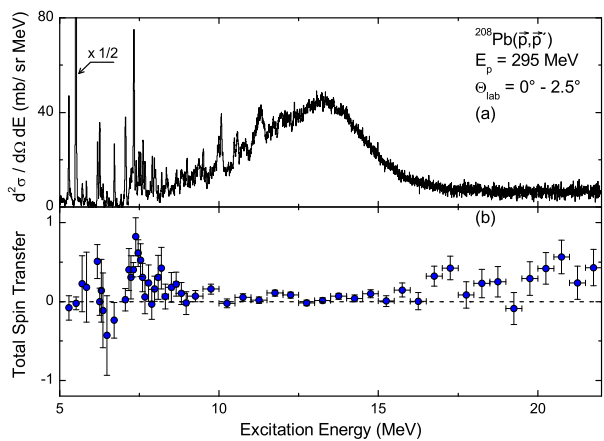}
\caption{\label{fig:spec}
(Color online) (a) Spectrum of the $^{208}$Pb($\vec{p},\vec{p}'$) reaction at $E_p = 295$ MeV with the spectrometer placed at $0^\circ$. (b) Total spin transfer $\Sigma$. }
\end{figure}

Figure~\ref{fig:spec}(a) displays a spectrum at $0^\circ$. Strong transitions at low excitation energies, a resonance-like structure close to $S_n = 7.37$ MeV and the prominent isovector giant dipole resonance (IVGDR), peaked at $E_x \approx 13.4$ MeV with pronounced fine structure, are observed.
The total spin transfer $\Sigma$ can be extracted from the measured PT observables which at $0^\circ$
%\begin{equation}
%  \Sigma = \frac{3-(2D_{SS'}+D_{LL'})}{4},
%  \label{eq:spintransfer}
%\end{equation}
takes a value of one for spinflip ($\Delta S = 1$) and zero for non-spinflip ($\Delta S = 0$) transitions \cite{suz00}.
Figure~\ref{fig:spec}(b) shows $\Sigma$ for $E_x = 5 - 22$ MeV. Values between 0 and 1 result from a summation over partially unresolved transitions with different spinflip character. The data reveal a concentration of spinflip strength in the energy region $ 7 - 9$ MeV, where the spin-$M1$ resonance in $^{208}$Pb is located \cite{hey10}, while the bump between 10 and 16 MeV has $\Delta S = 0$ character consistent with the excitation of the GDR.
% Shortened 24.5.11
The $\Delta S = 1$ strength above the GDR may result from the $\Delta S$ components of the onsetting quasifree scattering cross section \cite{bak97}.

A multipole decomposition was performed for angular distributions of the cross sections in the PDR and GDR regions.
Theoretical angular distributions were calculated with the code DWBA07 \cite{raynal} using microscopic quasiparticle-phonon model (QPM) wave functions \cite{rye02} and the Love-Franey effective proton-nucleus interaction \cite{fra85}. The interference of Coulomb and nuclear contributions to the cross sections was taken into account for $E1$ transitions. For a satisfactory description of the data it was sufficient to include, besides $E1$ and $M1$, one higher multipole representative for all other contributions. The latter was chosen to be either $E2$ or $E3$ in the region of the PDR. In the GDR region the $M1$ contribution was zero within error bars and was replaced by a phenomenological background describing the data at high excitation energies. The weight of each component was determined by a least-squares fit to the data.
%The examples in Fig.~\ref{fig:spec}(c) reveal a mixed $E1$/$M1$ composition of the energy bin around 7.3 MeV, while $E1$ dominates in the GDR region.

%
\begin{figure}[tbh]
\includegraphics[width=8cm]{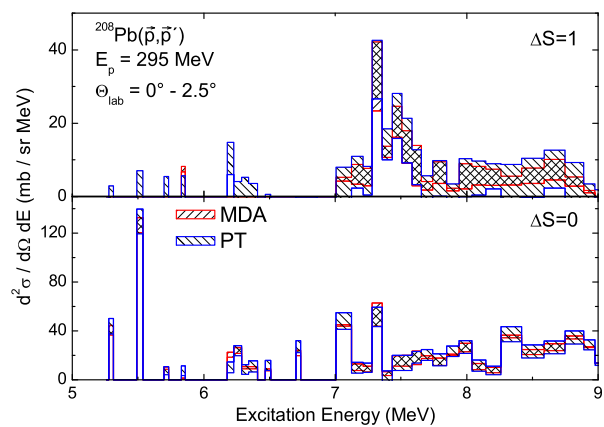}
\caption{\label{fig:compare}
(Color online) Decomposition of non-spinflip ($\Delta S = 0$) and spinflip ($\Delta S = 1$) cross section parts based on the MDA and PT, respectively, in the excitation energy region $5 - 9$ MeV. The hatched areas indicate the experimental uncertainties.
Excellent agreement between the two completely independent methods is observed.
}
\end{figure}
Cross sections for $\Delta S = 0$ and 1 from the MDA and PT analysis for $E_x < 9$~MeV are compared in Fig.~\ref{fig:compare}. Within uncertainties the correspondence between the two completely independent decomposition methods is excellent.
This puts confidence in the MDA results discussed in the following, which provide much better resolution because of the superior statistics  compared to a double scattering measurement of PT.
In the GDR region no direct comparison is possible because of the unknown $\Delta S$ content of the phenomenological background. However, both methods agree that $\Delta S =1 $ contributions are very small.
%
% Shortened 24.5.11
%This puts confidence in the MDA results discussed in the following, %which provide much better resolution because of the superior %statistics. In the GDR region, both methods agree that $\Delta S =1 $ %contributions are very small.

%
\begin{figure}[tbh]
\includegraphics[width=8cm]{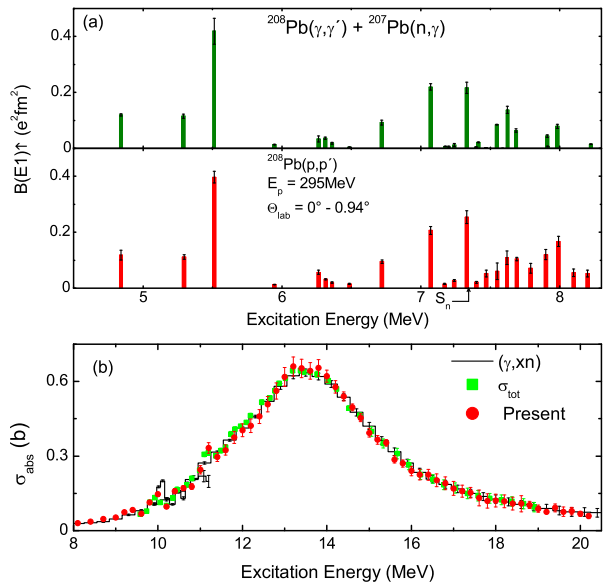}
\caption{\label{fig:be1comp}
(Color online) (a) B($E1$) strengths in $^{208}$Pb in the region $E_x \simeq 4.8 - 8.2$ MeV  as deduced from the present work in comparison with $(\gamma,\gamma')$ and $(n,\gamma)$ experiments \cite{rye02,end03,shi08,sch10}.
(b) Photoabsorption cross sections in the GDR region from the present work compared to $(\gamma,xn)$ \cite{vey70} and total photoabsorption \cite{sch88} measurements.
}
\end{figure}

Next we show that reliable B($E1$) strengths can be extracted from the $(p,p')$ data. While the angular dependence of $E1$ transitions is generally
state-dependent because of the Coulomb-nuclear interference,
cross sections at very small angles ($\Theta_{\rm lab} < 1^\circ$) arise purely from Coulomb excitation. Thus the conversion from cross section to strength is straightforward using semiclassical
theory \cite{ber88}. The B($E1$) distribution up to 8.2~MeV is compared in Fig.~\ref{fig:be1comp}(a) with an average over all available $^{208}$Pb$(\gamma,\gamma')$ and $^{207}$Pb$(n,\gamma)$ data (Refs.~\cite{rye02,end03,shi08,sch10} and refs.\ therein).
Excellent agreement is obtained up to $S_n$. The excess strength in the $(p,p')$ data above the neutron threshold can be attributed to previously unknown neutron decay widths of the excited $1^-$ states, which modify the branching ratios in  the $\gamma$-decay experiments and thus the extracted B$(E1)$ values. Figure~\ref{fig:be1comp}(b) shows the photoabsorption cross sections in the GDR region together with results from a $(\gamma,xn)$ \cite{vey70} and a total photoabsorption \cite{sch88} experiment. Again, very satisfactory agreement of all three measurements is observed.

\begin{figure}[tbh]
\includegraphics[width=8cm]{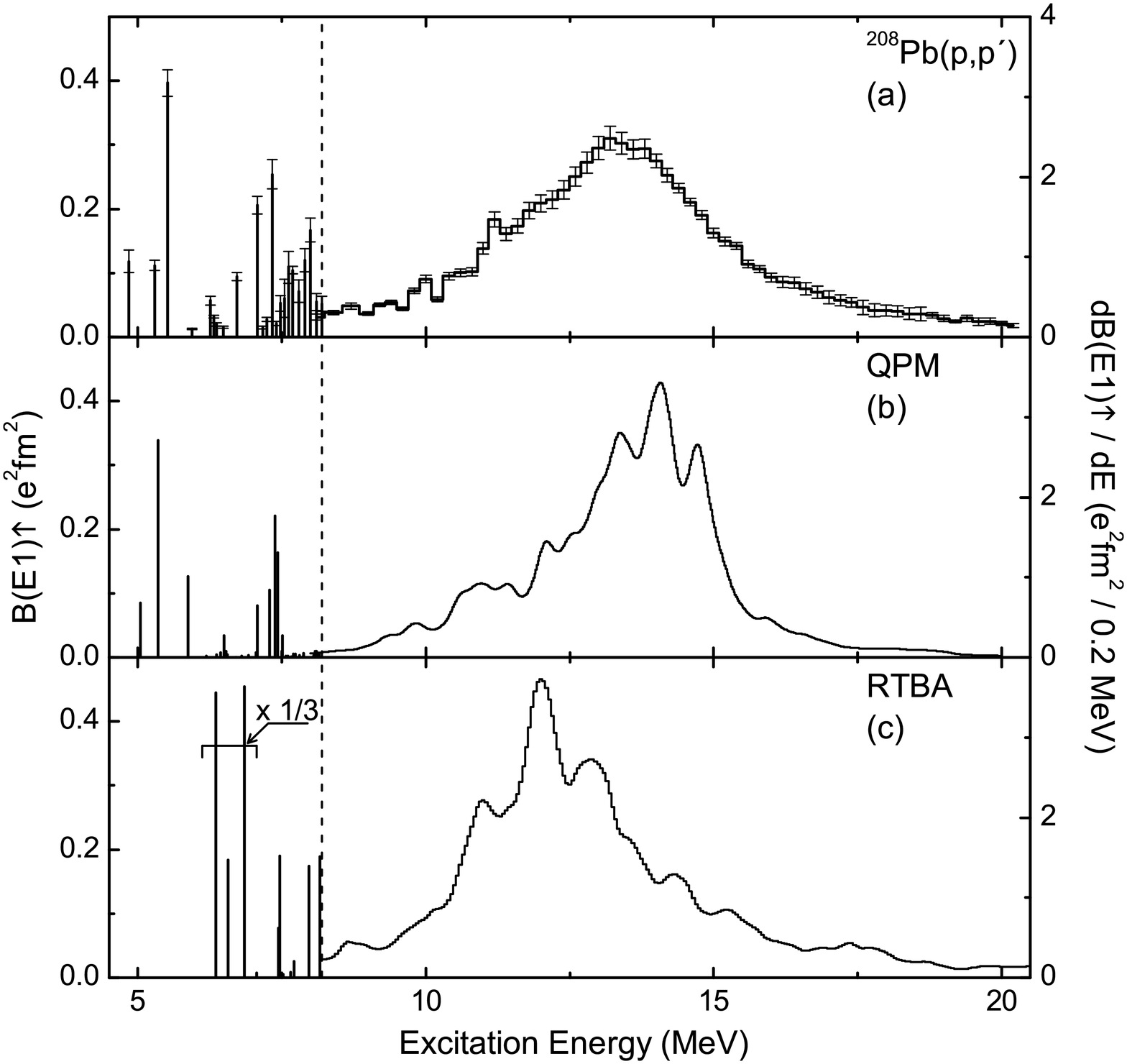}
\caption{\label{fig:be1expth}
Experimental B($E1$) strength distribution in $^{208}$Pb in comparison to QPM and RTBA calculations described in the text. Note the different scales below and above 8.2 MeV.
}
\end{figure}
Figure~\ref{fig:be1expth}(a) displays the experimental B$(E1)$ distribution. From the numerous computations of the $E1$ response in $^{208}$Pb we show in Fig.~\ref{fig:be1expth}(b) recent results from the QPM \cite{rye02}, and (c) the relativistic time-blocking approximation (RTBA) \cite{lit07}.
%Both methods allow for an inclusion of complex configurations.
The QPM calculations contain up to 3-phonon configurations for $E_x \leq 8.2$ MeV and 2-phonon configurations in the GDR region. Although the RTBA has recently been extended to include the full set of 2-phonon states \cite{lit10}, the results shown are based on a particle-hole$\otimes$phonon model space \cite{lit07}. In the low-energy region, the QPM provides a realistic description of the fragmentation but the overall strength is somewhat too small, while the RTBA model space is not yet sufficient to reproduce the fine structure, and the strength is somewhat too large. The width of the GDR is roughly reproduced by both models. Within the QPM the effective isovector interaction strength is adjusted to the experimental GDR centroid at 13.4 MeV. The RTBA calculations are fully self-consistent and the GDR centroid determined by the covariant EDF parametrization amounts to 12.9 MeV for the NL3 parameter set used.
%It remains unchanged when complex configurations beyond the mean-field level are included.
Such a comparison between high-precision data and the 3-phonon version
of the QPM guides the next generation of self-consistent extensions of
the covariant EDF.
% Shortened 24.5.11
Taking into account higher-order configurations, ground state correlations, and pairing vibrations should improve agreement with the data.

\begin{figure}[tbh]
\includegraphics[width=8cm]{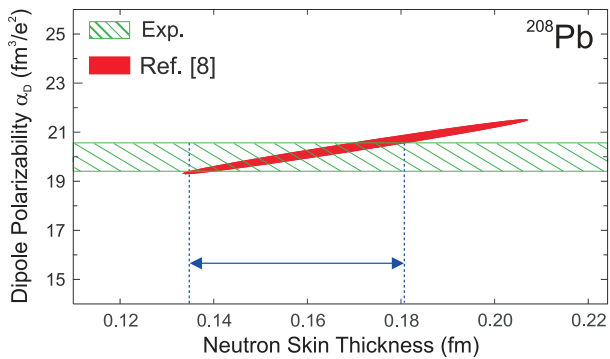}
\caption{\label{fig:polarizability}
(Color online) Extraction of the neutron skin in $^{208}$Pb based on the correlation between $r_{\rm skin}$ and the dipole polarizability $\alpha_D$ established in Ref.~\cite{rei10}.
}
\end{figure}

Finally, as discussed above, an important quantity is the electric dipole polarizability. We find $\alpha_D = 18.9(13)$ fm$^3$/$e^2$ for the $E1$ strength up to 20 MeV. By taking an average of all available data including excitation energies up to 130 MeV \cite{vey70,sch88}, a result with further reduced uncertainty $\alpha_D = 20.1(6)$ fm$^3$/$e^2$ is obtained. The covariance ellipsoid
%(cf.\ Fig.~1 of \cite{rei10})
of the correlation between $\alpha_D$ and the neutron skin thickness $r_{\rm skin}$ in the approach of Ref.~\cite{rei10} is shown in Fig.~\ref{fig:polarizability}. Only with the present precision for $\alpha_D$ (hatched band) one can constrain the neutron skin thickness to $r_{\rm skin} = 0.156^{+0.025}_{-0.021}$~fm. The hitherto most precise determinations of this quantity for $^{208}$Pb  \cite{fri07,zen10} deduced from exotic atoms ($r_{\rm skin} = 0.18\pm0.02$~fm) and hadron scattering ($r_{\rm skin} = 0.211_{-0.063}^{+0.054}$~fm), respectively, are in excellent agreement with our result based on a totally independent method.
%The corresponding QPM and RTBA results for $\alpha_D$($r_{\rm skin}$)   are 16.5~fm$^3$/$e^2$(0.155~fm) and 22.4~fm$^3$/$e^2$(0.288~fm), respectively.
Recent calculations of neutron matter and neutron star properties  \cite{heb10} in the framework of chiral effective field theory suggest $r_{\rm skin} = 0.17\pm0.03$~fm. The predictions are sensitive to three-nucleon forces, which may be further constrained by the present results. Since the correlation between polarizability, neutron skin thickness and symmetry energy is model-dependent, viz.\ $r_{\rm skin} \propto \alpha_D \cdot a_{\rm sym}$ \cite{sat06}, a systematic study with a variety of EDFs as well as experimental tests in other nuclei would be important.
%A combination of $\alpha_D$ with a model-independent determination of $r_{\rm skin}$ in $^{208}$Pb by the PREX experiment will allow an extraction of the density dependence of the symmetry energy $a_{\rm sym}$, since . Therefore, the present result provides an important constraint on the isovector properties of microscopic interactions.

To summarize, polarized proton scattering at very forward angles is a tool to study, with high resolution, the complete electric dipole response of nuclei from low excitation energies up to the GDR. The $E1$ strength distribution deduced in a benchmark experiment on $^{208}$Pb is in excellent agreement with available data. It provides, however, new information in the region around the neutron emission threshold where all previous experiments had limited accuracy. A precise value for the $E1$ polarizability can be extracted with important consequences for a determination of the neutron skin and the symmetry energy in neutron-rich nuclei. Although controversially discussed \cite{rei10}, $r_{\rm skin}$ may independently be derived from a similar correlation with the PDR strength \cite{pie10,kli07}, which is accurately determined by the present data as well.
Beyond these results, the experiment also confirms the spin-$M1$ resonance in $^{208}$Pb. Furthermore, the fine structure of the dipole modes contains information on level densities \cite{kal06} and characteristic scales \cite{she04}, giving insight into their dominant damping mechanisms.
%Systematic studies of the $E1$ and spin-$M1$ response in nuclei are currently underway with this new powerful technique.

We are indebted to the RCNP for providing excellent beams. Discussions with P.-G.~Reinhard and A.~Schwenk are appreciated. This work was supported by JSPS (Grant No.~14740154), DFG (contracts SFB 634 and 446 JAP 113/267/0-2). B.~R.\ acknowledges support by the JSPS-CSIC collaboration program and E.~L.\ by the LOEWE program of the State of Hesse (HIC for FAIR).

%\bibliography{pbprl}

\end{document}